\documentclass[aps,prl,twocolumn,epsfig,preprintnumbers,superscriptaddress,10pt]{revtex4}

\usepackage{graphicx}
\usepackage{dcolumn}
\usepackage{bm}

\newcommand{\be}{\begin{equation}}
\newcommand{\ee}{\end{equation}}
\newcommand{\bea}{\begin{eqnarray}}
\newcommand{\eea}{\end{eqnarray}}
\newcommand{\ba}{\begin{eqnarray}}
\newcommand{\ea}{\end{eqnarray}}

\newcommand{\eqn}[1]{(\ref{#1})}
\newcommand{\beq}{\begin{equation}}
\newcommand{\eeq}{\end{equation}}
\newcommand{\beqa}{\begin{eqnarray}}
\newcommand{\eeqa}{\end{eqnarray}}
\newcommand{\beqar}{\begin{eqnarray*}}
\newcommand{\eeqar}{\end{eqnarray*}}

\newcommand{\vlim}{v_\mt{lim}}
\newcommand{\ccbar}{c\bar{c}}
\newcommand{\ncc}{N_\mt{$c\bar{c}$}}
\newcommand{\opeak}{\omega_\mt{peak}}
\newcommand{\jpsi}{J/\psi}
\def\nc {N_\mt{c}}
\def\nf {N_\mt{f}}

\def\t7 {T_\mt{D7}}

\newcommand{\tf}{T_\mt{diss}}
\newcommand{\tc}{T_\mt{c}}

\newcommand{\mt}[1]{\textrm{\tiny #1}}

\begin{document}


\title{Prediction of a Photon Peak in Heavy Ion Collisions}

\author{Jorge Casalderrey-Solana} 
\affiliation{Nuclear Science Division, MS 70R319,
Lawrence Berkeley National Laboratory, Berkeley, CA 94720, USA}
\author{David Mateos}
\affiliation{Department of Physics,
University of California, Santa
Barbara, CA 93106-9530, USA}

\preprint{LBNL-72350}


\begin{abstract}
We show that if a flavour-less vector meson remains bound after deconfinement, and if its limiting velocity in the quark-gluon plasma is subluminal, then this meson produces a distinct peak in the spectrum of thermal photons emitted by the plasma. We also demonstrate that this effect is a universal property of all strongly coupled, large-$\nc$ plasmas with a gravity dual. For the $J/\Psi$ the corresponding peak lies between 3 and 5 GeV and could be observed in heavy-ion collisions at LHC. 
\end{abstract}

\maketitle

\noindent {\bf 1. Introduction.}
At a temperature $\tc \simeq 170$ MeV, Quantum Chromodynamics (QCD) undergoes a rapid cross-over into a deconfined phase called the quark-gluon plasma (QGP). This new state of matter is being intensively studied at the Relativistic Heavy Ion Collider (RHIC) and will be studied in the future at the Large Hadron Collider (LHC).

A remarkable conclusion from the RHIC experiments is that the QGP does not behave as a weakly coupled gas of quarks and gluons, but rather as a strongly coupled fluid. This makes the study of the plasma a challenging task. Experimentally, it is difficult to find clean probes with which to determine the properties of the plasma, since any coloured probe will strongly interact with the medium in a complicated manner; for this reason, thermal photons emitted by the plasma are particularly interesting. The theoretical task of predicting the properties of the QGP from first principles is also challenging. Since the plasma is strongly coupled, perturbative methods are not applicable in general. The lattice formulation of QCD is also of limited utility, since it is not well suited for studying real-time phenomena such as transport, photon production, etc. Thus one must either make predictions based on very general expectations, or resort to some other non-perturbative method such as the gauge/gravity duality. 

In this letter we will show that, under two plausible assumptions about the properties of heavy vector mesons in the QGP, a distinct peak in the spectrum of thermal photons is predicted. Moreover, we will demonstrate that this is a universal property of all strongly coupled, large-$\nc$ theories with a gravity dual. Finally, we discuss under what conditions this effect could be observed at LHC.

\noindent
{\bf 2. Peaks in the photon spectrum.}
Sufficiently heavy mesons (eg, $\jpsi, \Upsilon$, etc.) may be expected to exhibit two generic properties in the QGP. First, they may remain bound up to a dissociation temperature $\tf > \tc$. Second, their limiting velocity in the plasma may be subluminal. 

The original argument \cite{Matsui-Satz} for the first expectation is simply the fact that the heavier the meson, the smaller its size. It is thus plausible to expect a meson to remain bound until the screening length in the plasma becomes comparable to the meson size, and for sufficiently heavy mesons this happens at $\tf > \tc$.  This conclusion is supported by calculations of both the static quark-antiquark potential \cite{potential} and of Minkowski-space spectral functions in lattice-regularized QCD \cite{correlators}.

The second expectation goes back to ref.~\cite{Chu-Matsui}, which considered a weakly coupled plasma. More generally, one may note \cite{hotwind} that a meson moving in the plasma with velocity $v$ experiences a boosted, higher energy density, and hence also a higher effective temperature $T_\mt{eff}(v) = (1-v^2)^{-1/4} T$. Thus one may expect the existence of a subluminal limiting velocity for the meson \cite{alternative}, determined by the condition 
$T_\mt{eff}(\vlim) \sim \tf$ \cite{hotwind}.

Although the two assumptions above are reasonable, by no means they have been rigourously established in QCD. Our purpose is not to discuss their plausibility in detail but to exhibit an immediate consequence.

\begin{figure}
\includegraphics[scale=.2]{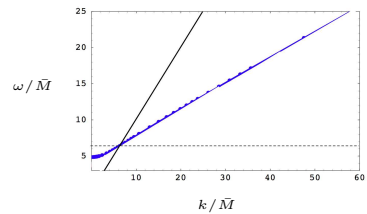}
\put(-135,35){$\mt{$\omega/\bar{M}$}$}
\put(-65,-10){$\mt{$k/\bar{M}$}$}
\caption{Dispersion relation (blue curve) for a heavy meson in the ${\cal N}=4$ SYM plasma at strong coupling \cite{MMT2}. The black straight line corresponds to 
$\omega=k$. $\bar{M}$ is a reference mass scale -- see \cite{MMT2} for details.} 
\label{dispersion}
\end{figure}

\begin{figure}[t!!]
\includegraphics[scale=.45]{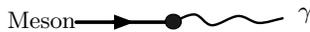}
\put(0,10){$\gamma$}
\put(-110,6){Meson}
\caption{Decay of a vector meson into and on-shell photon.} 
\label{decay}
\end{figure}

\begin{figure}[h!!]
\includegraphics[scale=.4]{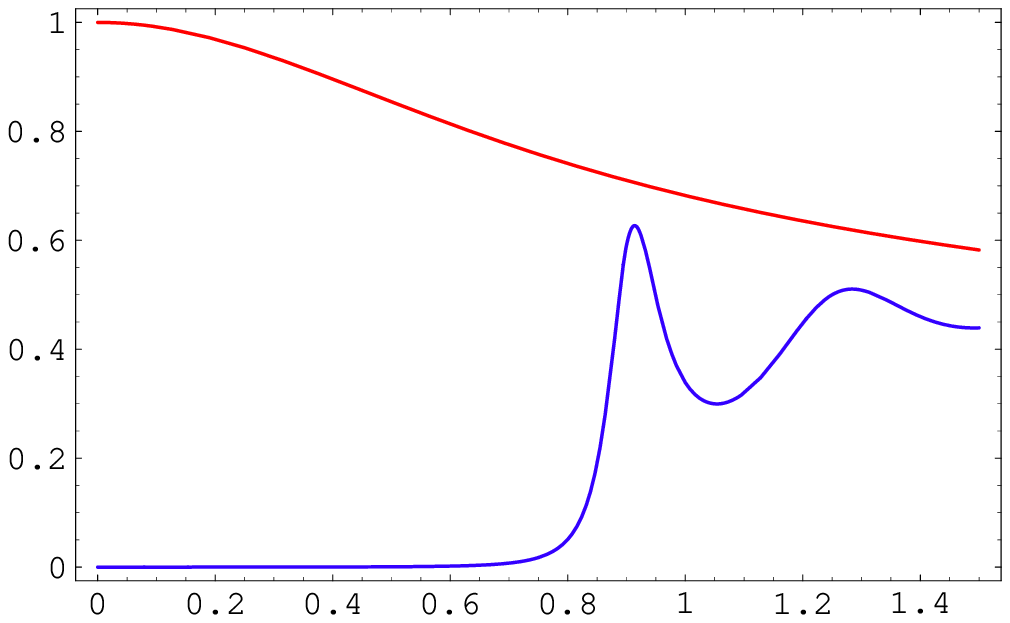} 
\put(-170,35){
{{$\mt{$\frac{\pi \chi^\mu_{\,\,\,\mu}  (\omega / 2 \pi T)}{\nc T \omega}$}$}}}
\put(-67,-10){$\mt{$\omega/2\pi T$}$}
\caption{Spectral functions for the ${\cal N} =4$ SYM plasma coupled to a massless quark (top, red curve) and a heavy quark (bottom, blue curve), at large $\nc$ and strong coupling.}
\label{illustration}
\end{figure}

Consider the in-medium meson dispersion relation $\omega(k)$, where $\omega$ and 
$k$ are the energy and the spatial three-momentum of the meson. $M=\omega(0)$ is the rest mass. As $k \rightarrow \infty$, the assumption of a subluminal limiting velocity implies $\omega(k) \sim \vlim k$, with $\vlim < 1$. Fig.~\ref{dispersion} shows the dispersion relation for a pseudoscalar meson in the strongly-coupled, large-$\nc$, four-dimensional, ${\cal N} =4$ super-Yang-Mills (SYM) plasma, calculated using its gravity dual \cite{MMT2}; the dispersion relation for vector mesons is expected to exhibit the same features. 
As noted in \cite{MP}, continuity now implies that the dispersion relation curve must cross the light-cone, defined by $\omega=k$, at some energy $\omega=\opeak$. At this point the meson four-momentum is null, and so the meson possesses the same quantum numbers as a photon 
\cite{same}. Such a meson can then decay into an on-shell photon \cite{on-shell}, as depicted in fig.~\ref{decay}. 
This process contributes a resonance peak, at an energy 
$\opeak$, to the in-medium spectral function of two electromagnetic currents, 
$\chi_{\mu\nu} (\omega,k) \sim \langle J_\mu (\omega,k) J_\nu(-\omega,-k) \rangle$, evaluated at null-momentum $\omega = k$. This in turn produces a peak in the spectrum of thermal photons emitted by the plasma, 
$dN_\gamma/d\omega \sim e^{-\omega/T} \, \chi^\mu_{\,\,\,\mu}  (\omega,T)$.
The width of this peak is the width of the meson in the plasma. 
In fig.~\ref{illustration} we have illustrated this effect for the ${\cal N}=4$ SYM plasma coupled to one massless quark and one heavy quark. The results are valid at strong coupling and large $\nc$, since they were obtained by means of the gravity dual 
\cite{MP}. The spectral function for the massless quark is structure-less, whereas that for the heavy quark exhibits a resonance peak -- see \cite{MP} for further details.

\noindent 
{\bf 3. A universal property of plasmas with a gravity dual.}
The gravity dual of QCD is presently unknown. When studying strongly-coupled plasmas with a gravity dual, it is therefore important to focus on properties that apply to as broad a class of plasmas as possible, since these may also apply to QCD. In this section we will show that the two assumptions above about heavy mesons in a QGP are true in all strongly coupled, large-$\nc$ plasmas with a gravity dual, because they follow from two universal properties of the duality: The fact that the deconfined phase is described by a  background with a black hole (BH) \cite{Witten}, and the fact that, in the large-$\nc$ limit, a finite number of flavours $\nf$ is described by $\nf$ D-brane probes in this background \cite{Karch-Randall}. 

\begin{figure}
\includegraphics[scale=.5]{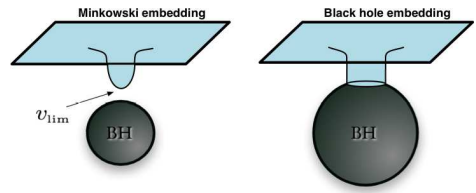}
\caption{Possible D-brane embeddings in a BH background.} \label{embeddings}
\end{figure}

In the presence of the black hole, there are two possible phases for the D-branes, separated by a universal first-order phase transition \cite{MMT,note}. Geometrically, these two phases are distinguished by whether or not the D-brane tension can compensate for the black hole gravitational attraction (see fig.~\ref{embeddings}).
In the first case the branes lie completely outside the horizon in a `Minkowski embedding'. In the second case they fall through the horizon in a `BH embedding'.  From the gauge theory viewpoint, this phase transition corresponds to the dissociation of heavy mesons 
\cite{MMT,Karl}. In the Minkowski phase stable mesons exist, and their spectrum is discrete and gapped. The meson mass in this phase increases as the separation between the branes and the black hole increases \cite{DDsystems}. By contrast, in the black hole phase no meson bound states exist. Recalling that the radius of the black hole is proportional to the plasma temperature, we see that if a meson is sufficiently heavy compared to the temperatrure, then this meson remains bound in the plasma and is described by a Minkowski brane. 
 
The existence of a subluminal limiting velocity for mesons is obvious from the geometric picture above: It is just the local speed of light at the tip of the branes \cite{MMT2}. Indeed, the wave function of a meson is supported on the D-branes. The larger the energy of the meson, the more it is attracted by the black hole and the more its wave-function is concentrated at the tip of the branes (see fig.~\ref{embeddings}). In the limit $k \rightarrow \infty$ the velocity of this meson approaches the local speed of light at the tip of the branes. Because of the redshift caused by the BH, this limiting velocity is lower than the speed of light at the boundary, where the gauge theory resides. In the gauge theory this translates into the statement that $\vlim$ is lower than the speed of light in the vacuum \cite{MMT2}. This effect is clearly illustrated in fig.~\ref{dispersion}.

\noindent
{\bf 4. Heavy Ion Collisions.}
Our analysis so far applies to an infinitely-extended plasma at constant temperature.  A crucial question is whether a peak in the photon spectrum could be observed in a heavy ion collision experiment. Natural heavy vector mesons to consider are the 
$J/\psi$ and the $\Upsilon$, since these are expected to survive deconfinement. We wish to compare the number of photons coming from these mesons to the number of photons coming from other sources. Accurately calculating the meson contribution would require a precise theoretical understanding of the dynamics of these mesons in the QGP, which at present is not available. Our goal will therefore be to estimate the order of magnitude of this effect with a simple model.

Following \cite{Rapp}, we model the fireball as an expanding cylinder with volume 
$V(t)=\pi(z_0+v_z t)(r_0 + a_\perp t^2/2)^2$ (we choose $t_\mt{thermalisation}=0$).
 This leads to the temperature evolution $T(t) = T(0) [V(0)/V(t)]^{1/3}$. At $t=t_\mt{hadro}$ the temperature reaches $\tc$ and the system hadronises. 

Let us first consider the $J/\psi$ contribution. In the spirit of statistical recombination 
\cite{statistical,Andronic:2006ky}, we assume that the $c\bar{c}$-pairs produced in the initial  collisions between partons become kinetically but not chemically equilibrated in the QGP 
\cite{Zhang:2008zzc}; in particular, their total number, $N_\mt{$c\bar{c}$}$, stays constant. To implement this condition we introduce a fugacity factor 
\cite{statistical,Andronic:2006ky}
\be
g_\mt{c}(T) = \frac{N_\mt{$c\bar{c}$}}{2 \cdot 3 \cdot V(T) \, 
\left( \frac{M_\mt{c} T}{2 \pi} \right)^{3/2} e^{-M_\mt{c}/T}} \,,
\ee
where $M_\mt{c}=1.7$ GeV is the charm-quark mass. At $t=t_\mt{diss}$ the temperature reaches the dissociation temperature of the $J/\psi$ in the medium, $\tf$. At this point a fraction of the $c\bar{c}$-pairs coalesce and recombine forming $J/\psi$ mesons \cite{subsequent}. Their contribution to the total number of photons with frequency $\omega$ emitted by the plasma is then
\be
S(\omega) \propto \int_{t_\mt{diss}}^{t_\mt{hadro}} dt \, V(t) \, g_\mt{c}(T(t))^2 \,  
e^{-\omega/T(t)} \, \chi_{\mt{$J/\psi$}} (\omega,T(t)) \,,
\label{signal}
\ee
where the `S' stands for `signal'. The overall normalisation is not important since it is the same as that for the background of thermal photons emitted by the light quarks, which is given by
\be
B(\omega) \propto \int_0^{t_\mt{hadro}} dt \, V(t) \, e^{-\omega/T(t)}  \,.
\label{B}
\ee
We have omitted $\chi_{\mt{light}}(\omega,T)$ since, guided by the results for plasmas with a gravity dual, we assume it is structureless. 

We would like to compare the two contributions above. Motivated by fig.~\ref{illustration}, we expect the magnitude of $\chi_{\mt{$J/\psi$}}$ around $\opeak$ to be comparable to that of $\chi_{\mt{light}}$. We thus model the spectral function for the $\jpsi$ by a unit-area Gaussian distribution of width $\Gamma$ centred at 
$\opeak(T(t))$.
For the width we choose $\Gamma = 100$ MeV, of the order of the temperature and a thousand times larger than the vacuum width. The temperature dependence of 
$\opeak$ arises from that of the meson dispersion relation. On general grounds we expect this to take the form 
$\omega(k) =  \sqrt{M^2 \vlim^4 + k^2 \vlim^2}+M (1-\vlim^2)$.
The first term is just the vacuum dispersion relation with the speed of light replaced by $\vlim$. The second term ensures that $M$ is the rest mass of the meson. The meson dispersion relation in fig.~\ref{dispersion} is very well approximated by this ansatz. We neglect medium-induced changes in the $\jpsi$ mass and set $M=3$ GeV. The above ansatz yields $\opeak = M ( 1+ \vlim/ \sqrt{1-\vlim^2})$. The limiting velocity is determined by the condition $(1-\vlim^2)^{1/4} = T / \tf + [ (1-v_0^2)^{1/4} -1 ]$.
The first term implements the relationship between $T$ and $\vlim$ that was heuristically motivated in sec.~2. The second term accounts for the fact that the limiting velocity at $\tf$ may be non-zero, as is the case in plasmas with a gravity dual. We choose $v_0 = 0.2$.
 
We also need the number of primordial $c\bar{c}$-pairs produced in the heavy ion collision. 
This is proportional to the $c\bar{c}$ cross-section in a pp-collision, which suffers from large uncertainties. For LHC energies we use $d\sigma_\mt{$c\bar{c}$}^{\mt{pp}}/dy = 1850$ $\mu$b, which is within the plausible range \cite{Cacciari}. With this choice one has $\ncc \simeq 60$. 

\begin{figure}
\includegraphics[scale=0.5]{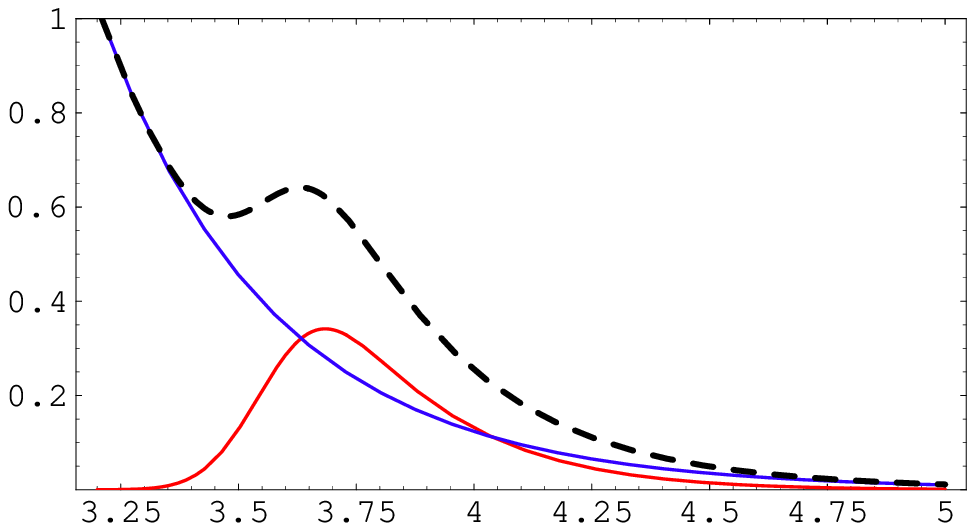} 
\put(-165,35){{\mt{$\frac{dN_\gamma}{d\omega}$}}}
\put(-75,-10){{\mt{$\omega$ [GeV]}}}
\caption{Thermal photon spectrum for LHC energies and $\tf=1.25 \tc$. The (arbitrary) normalisation is the same for all curves. The continuous, monotonically decreasing, blue curve is the background from light quarks. The continuous, red curve is the signal from $\jpsi$ mesons. The dashed, black curve is the sum of the two.}
\label{spectrum}
\end{figure}

The results of numerically evaluating eqs.~\eqn{signal} and \eqn{B} for LHC values of the parameters \cite{parameters} and $\tf=1.25 \tc$ \cite{mocsy} are plotted in fig. 
\ref{spectrum}.
We see that there is an order-one enhancement in the spectrum, associated to the decays of 
$\jpsi$'s into on-shell photons. However, whether this photon excess manifests itself as a peak, or only as an enhancement smoothly distributed over a broader range of frequencies, depends sensitively on the dissociation temperature. Perhaps surprisingly, the peak is less sharp for higher values of $\tf$. 

The signal is also quite sensitive to other parameters. For example, increasing the width by a factor of two turns the peak into an enhancement. Note also that $S$ depends quadratically on the $\ccbar$ cross-section. Since at RHIC  this is believed to be ten times smaller than at LHC, the enhancement discussed above is presumably unobservable at RHIC energies. Finally, we note that the signal is exponentially sensitive to other parameters such as the charm-quark mass, the maximum temperature $T(0)$, etc. 

These considerations show that a precise determination of the enhancement is not possible without a very detailed understanding of the in-medium dynamics of the $\jpsi$. On the other hand, they also illustrate that there exists a reasonable range of parameters for which this effect yields an order-one enhancement, or even a peak, in the spectrum of thermal photons produced by the QGP. This thermal excess is concentrated at photon energies roughly between 3 and 5 GeV. In this range the number of thermal photons at LHC is expected to be comparable or larger than that of pQCD photons produced in initial partonic collisions \cite{pQCD}. Thus we expect the thermal excess above to be observable even in the presence of the  pQCD background. 

We have examined the possibility of an analogous effect associated to the $\Upsilon$ meson. In this case $\opeak \gtrsim 10$ GeV. At these energies the number of thermal photons is exponentially smaller than that coming from pQCD photons, so  we do not expect an observable effect. 

\noindent
{\bf 5. Conclusions.}
Our two assumptions about the dynamics of vector mesons in the QGP can be motivated, and could in principle be checked, purely within QCD. We therefore emphasise that the prediction of a peak in the thermal photon spectrum does not rely on whether or not the QGP is well described by a dual gravity theory.

In practice, however, our ability to verify these assumptions in real-world QCD is very limited due to the lack of well-suited tools. It is therefore useful to investigate if they hold for strongly coupled, large-$\nc$ plasmas with a gravity dual, for which the gravity description provides such a tool. In this letter we have demonstrated that the two assumptions above are indeed true for all such plasmas.

\noindent 
{\bf Acknowledgments.} We thank M. Cacciari, D. d'Enterria, E. Ferreiro, S. Hartnoll, G. Martinez, V. Koch, M. Strassler, V.N. Tram and L. Yaffe for discussions. We are grateful to the INT at the U. of Washington for hospitality.
We are supported by the DOE, Contract No. DE-AC03-76SF00098 (JCS) and by the NSF grant PHY-0244764 (DM).


\begin{thebibliography}{99}
\bibitem{alternative}
An alternative possibility (not realised in plasmas with a gravity dual) would be that meson states with $k$ above some upper bound cease to exist. In any event, our conclusions will rely on meson states existing (with a sufficiently narrow width) only up to some moderate $k$ for which $\omega(k) \sim k$.
\bibitem{same}
We assume that the meson is flavour-less and we neglect the thermal mass of the photon.
\bibitem{on-shell}
In the vacuum, only the decay into a virtual photon would be allowed by kinematics.
\bibitem{note}
Specific examples of this transition in so-called Dp/Dq systems were originally seen in \cite{dpdq}. More recently similar transitions for Dp/Dq/D$\bar{\mbox{q}}$ systems \cite{SS} have been found in \cite{dpdqdqbar}.
\bibitem{DDsystems} 
In the Dp/Dq systems this is because the bare quark mass increases; in the Dp/Dq/D$\bar{\mbox{q}}$ systems this is because the chiral symmetry-breaking scale increases. In Dp/Dq systems at finite baryon density the branes always fall through the horizon, but for large quark masses the physics is essentially the same as at zero density \cite{finite}.
\end{thebibliography}
\end{document}